\documentclass[aps,twocolumn,showpacs,amsmath,prb,longbibliography]{revtex4-1}
\usepackage{graphicx}
\usepackage{subfigure}
\usepackage{amsmath}
\usepackage{bibentry}

\begin{document}

\title{Fano resonances can provide two criteria to distinguish Majorana
bound states from other candidates in experiments}

\author{Ye Xiong} 
\affiliation{Department of Physics and Institute of Theoretical Physics
  , Nanjing Normal University, Nanjing 210023,
P. R. China}
\affiliation{National Laboratory of Solid State Microstructures, Nanjing
  University, Nanjing 210093,
P. R. China}
\email{xiongye@njnu.edu.cn}

\begin{abstract}
  There are still debates on whether the observed zero
  energy peak in the experiment by Stevan {\it et al.} [Science 346,
  602(2014)] reveals the existence of the long pursuing Majorana bound
  states (MBS). we propose that, by mounting two scanning tunneling microscopic tips on
  top of the topological superconducting chain and measure the
  transmission spectrum between these two metallic tips,
  there are two kinds of characteristics on the spectrum that are caused
  by MBS uniquely. One is symmetric peaks with respect to zero energy
  and the other is $4\pi$ period caused by a nearby Josephson junction.
  The former refers to the fact that MBS are composited by Majorana
  fermions which distributed in the particle and hole subspaces equally.
  The latter is based on the well known $4\pi$ period of Josephson
  effect in topological superconductor. We think such two characteristics 
  can be used as criteria to distinguish MBS from other candidates, such
  as impurities, Kondo effect and traditional Andreev bound states.
\end{abstract}

\maketitle

\section{Introduction}\label{sec1}

Fano resonances (FR) \cite{PhysRev.124.1866} are a universal
interference phenomenon when waves coherently transfer through systems
with two kinds of paths. One is a media with continuous energy
spectrum and the other is scattering centers with discrete
eigen-energies. When the energy of an incoming wave is near those of
discrete levels, the total transmission fluctuates rapidly and forms an
asymmetric Fano line shape because the interference causes both resonant
and anti-resonant tunnelings \cite{RevModPhys.82.2257}. FR and
their line shapes are good experimental diagnosis on finding the nature
of transport paths\cite{PhysRev.124.1866}. The applications
include: tunneling measurement of Kondo scattering on single
magnetic atom attached on gold surface \cite{Madhavan24041998,
PhysRevLett.80.2893}, Raman spectrum of heavily doped semiconductors
\cite{PhysRev.158.748, PhysRevB.66.195206} and high-$T_c$
superconductors \cite{PhysRevLett.65.915, PhysRevB.61.4305}, electron
transport of quantum dots embedded in Aharonov-Bohm ring
\cite{PhysRevLett.88.256806, PhysRevB.68.235304} and carbon nanotubes
\cite{PhysRevLett.90.166403, PhysRevB.70.195408, PhysRevB.70.153405},
conductance of quasi-one-dimensional quantum wires with donor impurities
or quantum dots \cite{PhysRevB.48.2553, PhysRevB.67.155301}, and so on.
External parameters can also modify FR. For instance, the external bias
can change the asymmetric absorption spectrum of the Wannier-Stark
transitions in semiconductor superlattices \cite{PhysRevB.71.205326,
0295-5075-74-5-875}. Besides these, dynamical FR takes place when the Rabi
oscillations and coherent phonons interfere, in which the Rabi
oscillations with broadened linewidth play the role of continuous
spectrum \cite{PhysRevLett.115.157402}. Because FR provides an accurate
method to measure the linewidth of discrete levels, it was also proposed
to link FR with the measurement of phase decoherence in mesoscopic
system \cite{0295-5075-73-2-164}. 

In this paper, we propose to setup a feasible experiment and use FR
to answer the debated question on whether the experimentally observed zero energy
modes are associated with the Majorana bound states (MBS). MBS are Majorana
fermions in condensed matter physics, which are their own antiparticles
\cite{1063-7869-44-10S-S29, 0034-4885-75-7-076501, RevModPhys.80.1083}.
Recently, these long-pursuing quasiparticles are claimed to be observed
experimentally in various systems, including nanowire-superconductor
junctions \cite{doi:10.1021/nl303758w, Das2012b,
PhysRevLett.110.126406}, InSb/Nb junction \cite{Rokhinson2012}, chains
of magnetic atoms on a surface of superconductor
\cite{Nadj-Perge31102014}, and vortex
in $Bi_2Te_3 / NbSe_2$ heterostructure \cite{PhysRevLett.114.017001}.
However, alternative explanations based on impurities, Kondo effect or
Andreev bound states (ABS) on a ferromagnetic wire have been proposed
\cite{PhysRevLett.109.267002, 1367-2630-14-12-125011,
PhysRevLett.109.227005, PhysRevB.86.100503, PhysRevLett.115.127003}.
These criticisms mostly stem from the fact that the experiments only
measure the current through a scanning tunneling microscopy (STM) tip,
or through the heterostructure.
The observed zero bias peaks only reveal a finite local density of
states at zero energy. But such peaks can not be associated with MBS
uniquely. 

Our proposal is based on the fact that MBS at the zero energy are the
quantum states composited by half fermions. As a result, their weights
in the particle and hole subspaces in the Nambu representation are
balanced. But for other Bogoliubov quasi-particles, their weights in the
particle subspace are $|u|^2= \frac{1}{2} (1+\xi/E)$ and
$|v|^2=\frac{1}{2} (1-\xi/E)$ in the hole subspace, where
$E=\sqrt{\xi^2+\Delta^2}$ is the eigen-energy. These weights in the two
subspaces are not equal in general. As we will show, FR can identify
whether such balance is broken or not. By this way, it is easy to
distinguish the effect of MBS from impurities or Kondo effect. But such criterion may
not be sufficient to distinguish the MBS from the near-zero-energy ABS, as the
latter may also have nearly balanced weights in the two subspaces. We
then propose another criterion which is based on the well known $4\pi$
period of the current through a Josephson junction in the topological
superconductor \cite{PhysRevLett.105.077001, PhysRevB.79.161408}. We
will show that although finite size effect induced anti-crossing
destroys the $4\pi$ period in the energy spectrum, FR can still reveal
this double period. In such a way, we may finally distinguish MBS from a
traditional ABS in experiments. At the end of this paragraph, we want to
emphasize that this proposal is mostly suitable to be applied in the
system of magnetic atoms on superconductor \cite{Nadj-Perge31102014} and
we have not posed any new experimental challenges.

\section{Distinguishing MBS with Fano resonances}

We propose to place two metallic STM tips on top of the left end of a
topological superconducting (TS) chain possessing MBS. When the bias
voltage between the STM tips is smaller than the superconducting gap of
the underlying topological superconductor, FR should be observed on the
transmission spectrum between the two STM tips. Here MBS are taken as
the discrete levels near the zero energy and the metallic STM tips provide
the necessary continuous energy spectrum. The superconductor beneath the TS
chain is separated into two parts so that a Josephson junction is formed
in the above TS chain. The phase of superconducting order in the two
parts can be varied temporally by an external voltage. 

\begin{figure}[ht]
  \centering
  \includegraphics{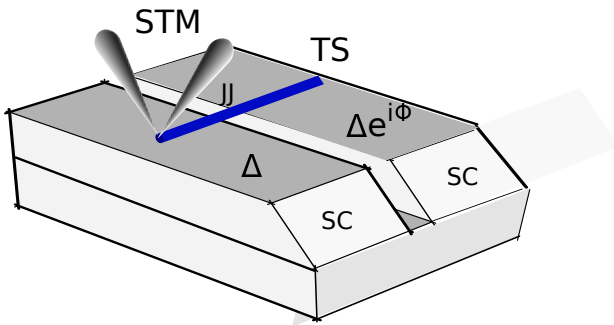}
  \caption{A sketch of the proposed experimental setup. The blue
  thick line is the TS chain placed on top of superconductor (SC).
  The two parts of superconductor are different by a phase, $\Phi$, in
  the order parameter $\Delta$. Such phase can be induced by an external voltage 
  or a magnetic flux not shown here
  explicitly. An ABS whose energy is proportional to $\cos(\Phi/2)$ is
  created in the TS chain at the Josephson junction (JJ). Two metallic
  STM tips are placed on top of the left end of the TS chain. In
  experiments, one measures the transmission spectrum between these
tips.}
  \label{fig1}
\end{figure}

As Fig. \ref{fig1} shows, we consider a TS chain coupled with two STM
tips. The Hamiltonian reads
\begin{equation}
  H=H_{K}+H_{L}+H_{R}+H_{\text{hyb}},
  \label{eq1}
\end{equation}
where we use the spinless Kitaev model on discrete lattice to
simulate the TS chain:
\begin{eqnarray}
  H_K & = &\sum_{i=1}^{N} \mu c^\dagger_{i}c_i + \sum_{i=1}^{N/2} (t
  c^\dagger_ic_{i+1} + \Delta c^\dagger_i c^\dagger_{i+1} +\text{h.c.}
  ) \nonumber \\
   & & +  \sum_{i=N/2}^{N-1} (t
  c^\dagger_ic_{i+1} + \Delta e^{i\Phi} c^\dagger_i c^\dagger_{i+1} +\text{h.c.}
  ).  
  \label{eq2}
\end{eqnarray}
Here, $N$ is the total number of sites, $\mu$ is on-site energy, $t$ is
the strength of hopping between the nearest neighboring sites, and $\Delta$
is the strength of the p-wave superconducting pairing. 
$\Phi$ is the phase difference inherited from that of the beneath
superconductors. $c^\dagger$ and $c$ are fermionic creation and
annihilation operators.

The spinless Kitaev model is plausible to catch the major physics of a
TS chain, as among the two spin degree of freedom, only one effectively
contributes to the topological nontrivial bands and the other has
already been gapped out by, typically spin-orbital interaction. Our
setup of TS chain is similar to those proposals for the measurements of
the $4\pi$ periodic Josephson current \cite{PhysRevLett.105.077001,
PhysRevB.79.161408, PhysRevB.84.081304, PhysRevLett.108.257001}. But for
magnetic atoms on top of superconductor, such longitudinal current
measurement is hard to perform in experiments. We also ignore the
decoration terms, such as gate or smaller hopping across the junction,
as these terms will not alter the spectrum of the ABS qualitatively,
that is $E \sim \pm \cos(\Phi/2)$, at the junction. 

\begin{equation}
  H_{L(R)} = \sum_{i=0}^\infty ( t_0 d^\dagger_{i,L(R)} d_{i+1,L(R)}
  +\text{h.c.} )
  \label{eq3}
\end{equation}
are the two semi-infinite metallic chains stand for the two STM tips.
$2t_0$ is their band width and $d^\dagger$ and $d$ are the creation and
annihilation operators on these tips. For the sake of clarify, we only
take one channel in each chain. Introducing multi-channels in each tip will
induce nothing but increase of the total conductance by a factor.

The Hamiltonian that hybridizes the TS chain and the two STM tips reads
\begin{eqnarray}
  H_{\text{hyb}} & = &  t_{\text int} (d^\dagger_{0,L} d_{0,R} +\text{
  h.c.}) \nonumber \\
  & & +\sum_{\alpha=L,R} \sum_{i=0}^M (J_i c^\dagger_1 d_{i,\alpha}
  +\text{h.c.}).
  \label{eq4}
\end{eqnarray}
Here the first term represents the hopping between the two STM tips and
the second one stands for the hopping between the TS chain and the STM
tips. $M$ is the range out of which such hoppings are ignored.

\begin{figure}[htp]
  \begin{minipage}[t]{0.5\textwidth}
  \centering
  \includegraphics[width=0.45\textwidth]{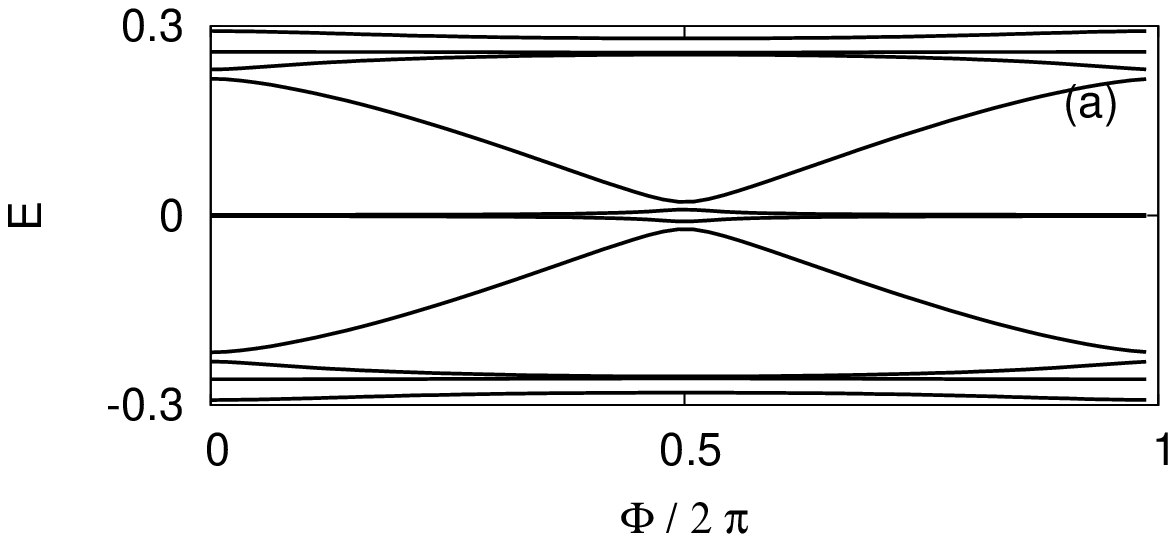}
  \includegraphics[width=0.45\textwidth]{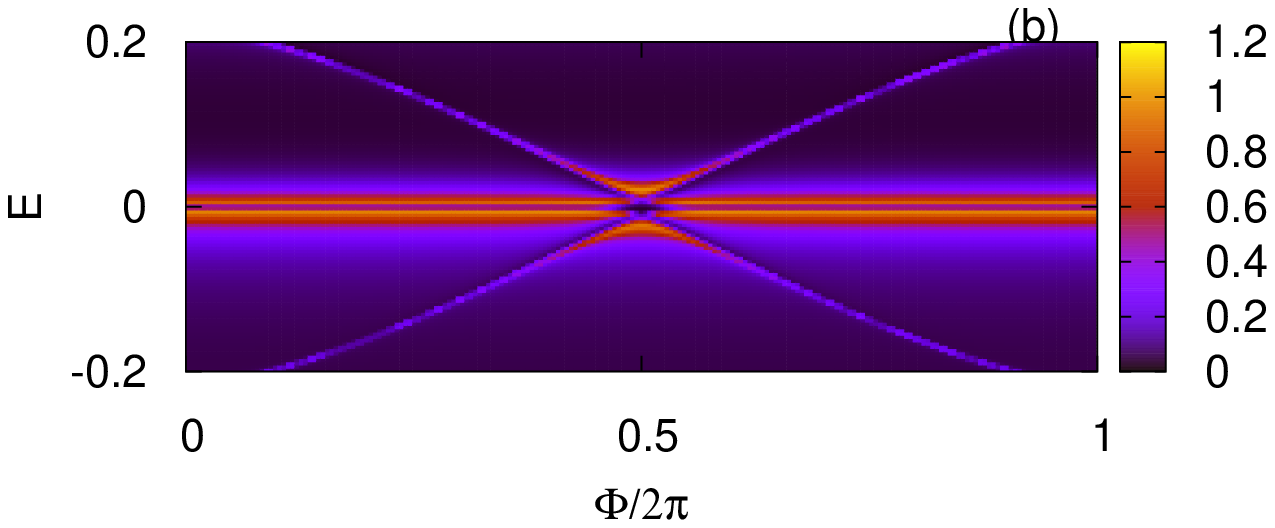}
  \end{minipage}
  \begin{minipage}[t]{0.5\textwidth}
  \centering
  \includegraphics[width=0.45\textwidth]{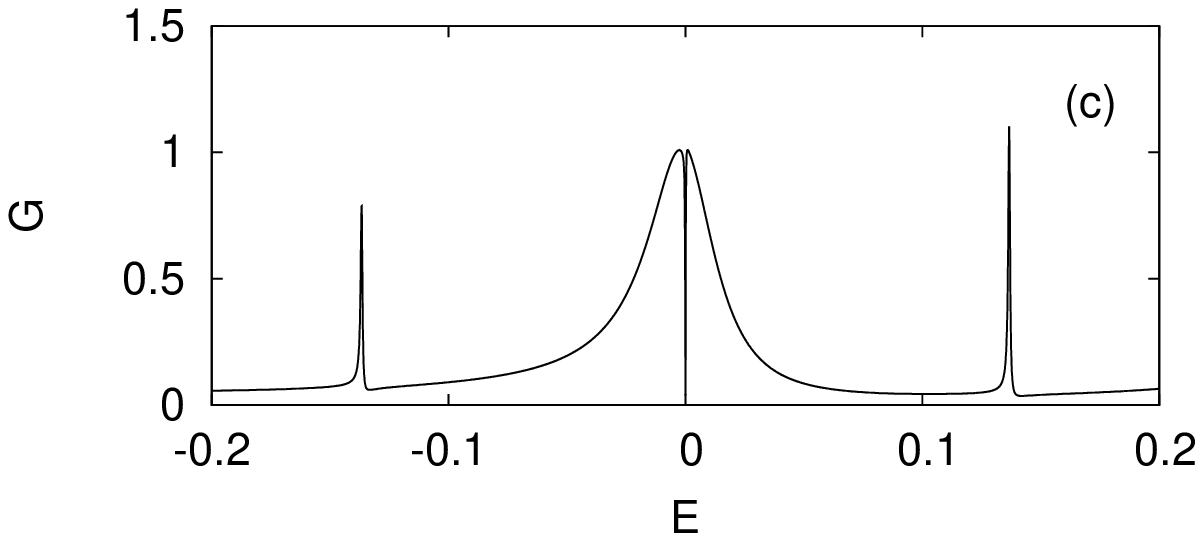}
  \includegraphics[width=0.45\textwidth]{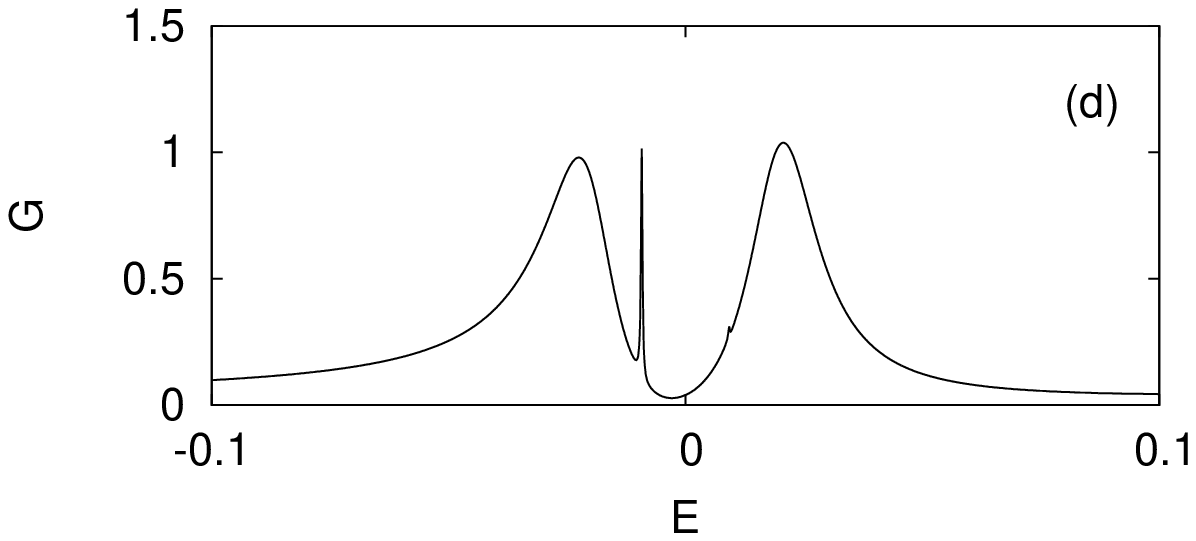}
  \end{minipage}
  \caption{(a) The energy spectrum of an isolated TS chain. Here
  $\mu=0$, $t=0.2$, $\Delta=0.1$ and $N=10$. In the superconducting
  gap, beside the MBS at the zero energy, there is also an ABS localized
  at the junction. Its energy is varying with $\Phi$ as $\cos(\Phi/2)$.
  A finite size effect induces an anti-crossing at $\Phi=\pi$. (b)
  Transmission spectrum with respect to $\Phi$ and energy $E$. (c) and
  (d) the detailed transmission spectrum v.s. $E$ at $\Phi=\pi/2$ and
  $\pi$, respectively. 
}
  \label{fig2}
\end{figure}

In Fig. \ref{fig2}(a), we present the spectrum of an isolated TS
chain. Such chain is within TS phase so that there is an MBS at the zero
energy (although finite size effect lifts the energy a little from
exact zero). Beside MBS, the fractional Josephson effect in TS chain
induces an ABS whose energy is varying with $\Phi$ as $\cos(\Phi/2)$.
This is why the ABS state at the Josephson junction is approaching
zero energy at $\Phi=\pi$, instead of $\pi/2$ in traditional
superconductors. Around zero energy, finite size effect hybridizes it
with MBS and induces the small anti-crossing there. 

We then mount the two metallic semi-infinite chains on top of an end of the TS
chain, The electric transmission through these metallic chains is
characterized by the differential conductance $G$ (in the units of
$\frac{e^2}{h}$) shown in Figs.
\ref{fig2}(b), (c) and (d). Here the parameters are taken as $t_0=1$,
$t_{int}=0.1$, $J_i=0.1$ and $M=5$. The conductance $G$ is calculated by
the BTK method \cite{PhysRevB.25.4515} which is explained in details in Appendix. 

Figures \ref{fig2}(c) and (d) are for the cases with $\Phi=0.5\pi$ and
$\pi$, respectively. In Fig. \ref{fig2}(c), the broadened central peakes
are caused by the resonant transmission though the MBS at the zero
energy. We notice that there is a sharp dip at the zero energy that
splits the center peak into two peaks. Such splitting can be understood
as the finite-size caused level splitting for MBS. We also notice that
the heights of the two peaks are identical and their shapes are
near-symmetric with respect to the zero energy. After comparing this FR with
that caused by an impurity state, we will show that such near
symmetric peaks around zero energy cannot appear in the impurity case.
There are also small, sharp peaks decorated on the board peak at $|E|
\sim 0.13$. These
small peaks are cause by the resonant tunneling through ABS at the
nearby Josephson junction. One notices that the heights of these two
peaks are different. This asymmetry stems from the fact that the
eigenstate of ABS is asymmetric in particle and hole space. In Fig.
\ref{fig2}(d), $\Phi=\pi$ and the anti-crossing takes place rightly. So
the two broad peaks are pushed away from zero energy and the remaining
sharp peak near zero energy is caused by the ABS. 

In Fig. \ref{fig2}(b) we show how these peaks evolute with $\Phi$.
Figures \ref{fig2}(c) and (d) are the two slides taken from it at $\Phi=\pi/2$
and $\pi$, respectively. 

\begin{figure}[htp]
  \begin{minipage}[t]{0.5\textwidth}
  \centering
  \includegraphics[width=0.45\textwidth]{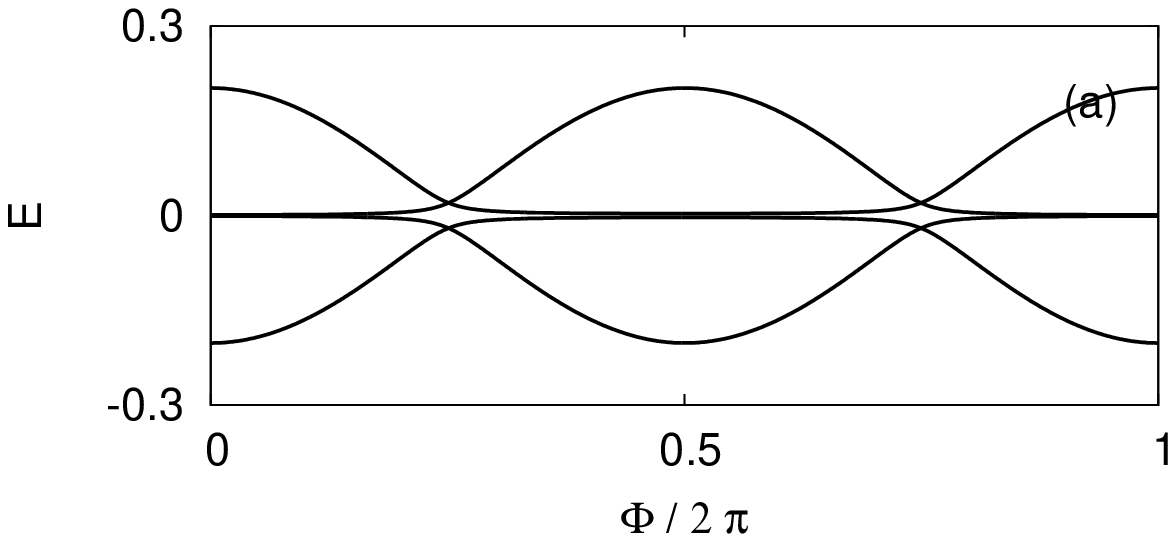}
  \includegraphics[width=0.45\textwidth]{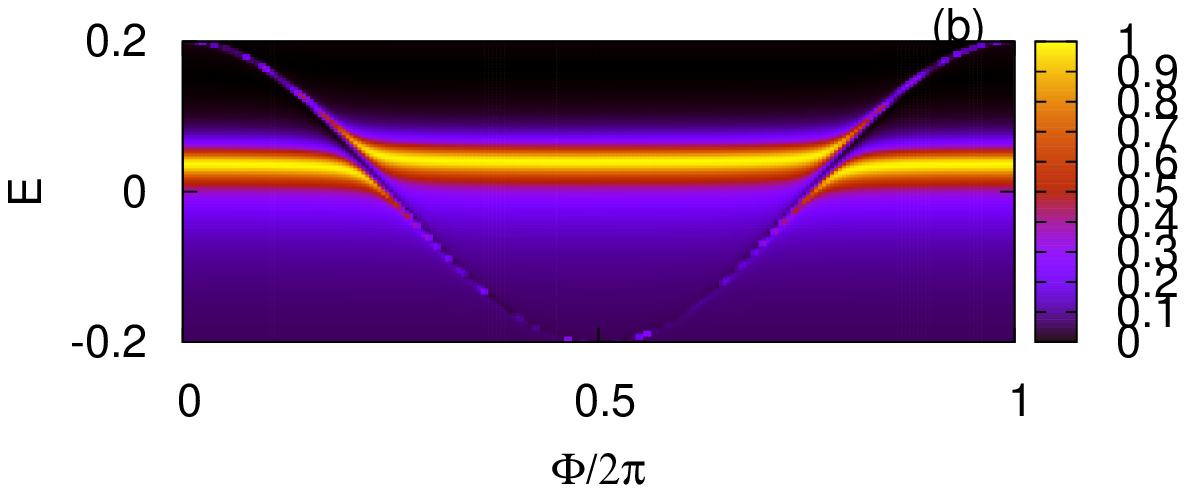}
  \end{minipage}
  \begin{minipage}[t]{0.5\textwidth}
  \centering
  \includegraphics[width=0.45\textwidth]{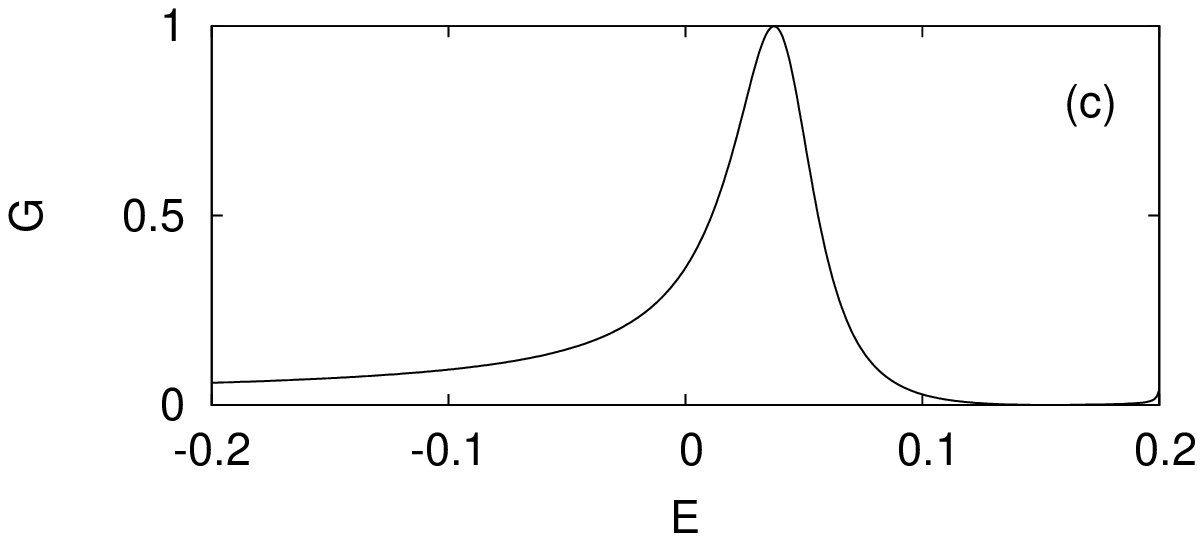}
  \includegraphics[width=0.45\textwidth]{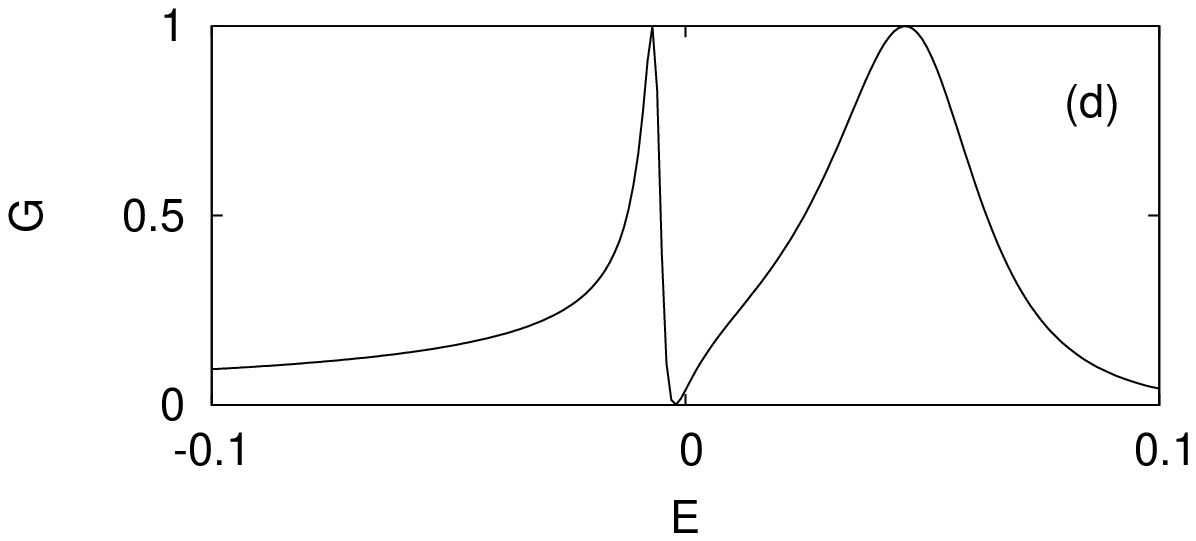}
  \end{minipage}
  \caption{(a) The energy spectrum of an isolated scattering center
  composited by impurities. Here $\varepsilon_0=0.001$,
  $\varepsilon_1=0.2$, $t_1=0.02$ and $\Delta_1=0$. Nambu
  representation is taken so that there is particle-hole symmetry with
  respect to $E=0$.(b)
  Transmission spectrum with respect to $\Phi$ and energy $E$. (c) and
  (d) the detailed transmission spectrum v.s. $E$ at $\Phi=0$ and
  $\pi/2$, respectively. 
}
  \label{fig3}
\end{figure}

To explicitly show the unique characteristics of FR of MBS, we plot what kind
of resonant peaks they will look like when the scattering center is made
from impurities.
Such system can be described by a similar Hamiltonian as Eq. \ref{eq1}
with $H_K$ replaced with
\begin{eqnarray}
  H_{\text imp} & = & \varepsilon_0 c^\dagger_0 c_0 + \varepsilon_1
  \cos(\Phi) c^\dagger_1 c_1 \nonumber \\
  & & +t_1 c^\dagger_0 c_1 +\Delta_1 c^\dagger_0 c^\dagger_1.
  \label{eq5}
\end{eqnarray}
Here we simplify the scattering center with only two effective states
and ignore all other states out of the gap. $c_1$ and $c_2$ are the
annihilation operators on these states respectively. State $1$ is
representing the impurity state at the end of the chain accidentally
having energy $\varepsilon_0 \sim 0$ and state $2$ stands for the possible
ABS at the junction. The energy of ABS should be a periodic function of
$\Phi$, which is written down as $\varepsilon_1 \sim \cos(\Phi)$. The
strength of hopping between the two states is $t_1$. For the sake of
clarify, we take $\Delta_1=0$ in Fig. \ref{fig3}. Such approximation
runs to an extreme that in the Nambu representation, the states above
the zero energy are particle-like with $u=1$ and those states below the
zero energy are hole-like with $v=1$. We take such extreme to
illustrate that the transmission in our setup is only contributed by
the particle subspace in the Nambu representation.

In Fig. \ref{fig3}(a), we plot the energy spectrum of the isolated
scattering center. Nambu representation is taken although $\Delta_1=0$.
In Fig. \ref{fig3}(c) and (d) we plot the differential conductance
through the metallic leads when $\Phi=0$ and $\pi/2$ respectively. We
find that only the particle-part of the impurity states contribute to
the conductance, while their imaging states in the hole-subspace are
invisible in the transmission spectrum. This is easy to be understood in
such an extreme case. As the superconducting order $\Delta_1$ is zero,
Nambu representation will introduce an artificial trivial duplication.
These FR are well understood when we give up the Nambu representation
and only consider the model within the particle-subspace.

\begin{figure}[htp]
  \centering
  \includegraphics[width=0.45\textwidth]{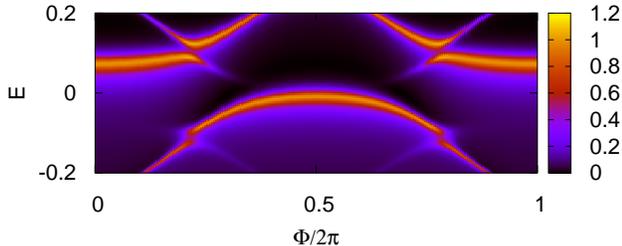}
  \caption{
  Transmission spectrum with respect to $\Phi$ and energy $E$. Other
  parameters are the same as those in Fig. \ref{fig3} except 
  $\Delta_1=0.1$.
}
  \label{fig4}
\end{figure}

Such extreme model indicates that only the particle part of the impurity
states will contribute to the FR. This is further confirmed when we
switch on $\Delta_1$. In Fig. \ref{fig4}, we show the differential
conductance with $\Delta_1=0.1$. The peaks at $E>0$ and $E<0$ are highly
asymmetric. This is because except MBS, other states in superconducting
gap, are generally unbalanced in the  particle and hole sub-spaces. 

After compare the FR of the two models, we conclude that the FR  can
provide two criteria for the verification of MBS in experiments. One is
the symmetric broadened peaks around zero energy when $\Phi$ is around
$0$. So such criterion is still when the Josephson junction is
eliminated. It can exclude the possible impurity effect or the Kondo
effect solidly from the candidates. But such criterion may not
distinguish the MBS from the near-zero-energy ABS,
 because both of them have almost equal weights in the two
subspaces. 

In this case, we need the second criterion, which must engage a
Josephson junction as we have shown in Fig. \ref{fig1}.  A traditional
ABS must evolve with $\Phi$ periodically, while it because $4\pi$ for
MBS. As a result, we can exam when and how much time the FR peaks cross
(or anti-cross) the zero energy as varying $\Phi$ from $0$ to $2\pi$. 
As shown in Fig. \ref{fig2}(b), in topological superconductor, the cross
(or anti-cross) takes place at $\Phi=\pi$ and only takes place one time
in a period $\Phi=[0,2\pi]$. While in traditional superconductor with
ABS, as shown in Fig. \ref{fig3}(b), such cross (or anti-cross) takes
place at $\Phi=\frac{\pi}{2}$ and $\frac{3\pi}{2}$, and appears twice in
one period.

With these two criteria, we can undoublly exclude the impurity states,
the Kondo effect and the ABS from the candidates of the
possible explanations. They may help us go a little further on the route
to MBS. Although the compared model shown in Fig. \ref{fig3} seems tricky and
crude, it catches the major physics we are facing with.
The range of energy we are interested in is within the superconducting
gap. This allows us simplify the scattering center with only two states.
We do approximate these two states as regular fermion states, while the
in-gap states should be Bogoliubov states with both components in the
particle and hole subspaces. But as soon as their energies are not fixed
at zero, their components in these two subspaces are unbalanced. As a
result, our first criterion about the absence of the symmetry of Fano
peaks with respect to zero energy is not depended on the approximation. One can
further confirms that in Fig. \ref{fig2}(d), as the heights of the two
sharp peaks caused by the ABS are not equal. 

\section{Conclusions}

We have proposed an experiment-friendly setup to verify the existence of
MBS in experiments. Unlike the previous proposals which require
a relative long chain to eliminate the finite size effect, this proposal
takes benefits from the finite size effect so that only a shorter TS
chain is required. By measuring the transmission spectrum through STM
tips mounted on top of TS chain, one can distinguish MBS from
impurities, kondo effect. One can also watch the fractional Josephson
effect from the transmission spectrum after introducing a Josephson
junction in the chain. Under these judgments, the experimentalists may
undoubtedly claim the observation of MBS in experiments.

After completing this paper, we also notice the experimental and
theoretical works on the transport characteristics of a TS chain with
only one STM superconducting tip on top\cite{PhysRevLett.115.197204,
Peng2015b}. It seems that MBS will also induce balance peaks at the edge
of tip's gap in that case. 

\appendix

\section{BTK method}

BTK method is used to calculate the scattering matrix though the
metal-superconductor-metal interfaces. As only one channel is taken in
each metallic lead, we can write down the incoming wave, the
reflected wave and the transmitted wave in the Nambu representation as
\begin{eqnarray}
  \psi_{in} &=& \begin{pmatrix} 1 \\ 0 \end{pmatrix} e^{ik_p x} \\ 
  \psi_{refl} &=& A \begin{pmatrix} 0 \\ 1 \end{pmatrix} e^{ik_h x} + B
  \begin{pmatrix} 1 \\ 0 \end{pmatrix} e^{-ik_p x} \\
  \psi_{trans} &=& C \begin{pmatrix} 0 \\ 1 \end{pmatrix} e^{-ik_h x} + D
  \begin{pmatrix} 1 \\ 0 \end{pmatrix} e^{ik_p x}, 
  \label{}
\end{eqnarray}
where $k_p$ and $k_h$ is the wave vector of particle and hole, with
their group velocities pointing to the correct directions during this
scattering process. Here $2t_0\cos(k_p)=-2t_0 \cos(k_h) =E$. The total 
wave-functions are $\psi_{in}+\psi_{refl}$ in the left lead and 
$\psi_{trans}$ in the right lead. We then solve the Schrödinger 
equation to find these parameters $A$, $B$, $C$ and $D$. The 
transmission spectrum is calculated by
\begin{equation}
  G= \frac{e^2}{h} (1-|B|^2+|A|^2). 
  \label{}
\end{equation}

\bibliographystyle{apsrev}
\bibliography{fano}

\end{document}